# Performance Considerations
# Gigabyte per Second Transcontinental Disk-to-Disk File Transfers

Peter Kukol and Jim Gray

Microsoft Research

# Performance Considerations
# Gigabyte per Second Transcontinental Disk-to-Disk File Transfers

Peter Kukol, Jim Gray, Microsoft Research
9 July 2004

**Abstract:** Moving data from CERN to Pasadena at a gigabyte per second using the next generation Internet requires good networking and good disk IO. Ten Gbps Ethernet and OC192 links are in place, so now it is simply a matter of programming. This report describes our preliminary work and measurements in configuring the disk subsystem for this effort. Using 24 SATA disks at each endpoint we are able to locally read and write an NTFS volume is striped across 24 disks at 1.2 GBps. A 32-disk stripe delivers 1.7 GBps. Experiments on higher performance and higher-capacity systems deliver up to 3.5 GBps.

**Summary**: We've been working with Cal Tech (Yang Xia, Harvey Newman, et al) and CERN (Sylvain Ravot, et al) to move data between CERN and Pasadena at 1GBps using the Internet rather than sneaker net. Our networking colleagues (Ahmed Talat, Inder Sethi, et. al.) have a good start on using 10 Gbps Ethernet to move 1GBps across the planet (ultralight). We (Kukol and Gray) are working on the first-meter last-meter problem of quickly moving data from disk to NIC and NIC to disk.

To do that we need roughly 1.2 GBps of disk I/O bandwidth (a 20% margin allows us some slack.) That translates to about 20 disk drives at the outer band (60GBps/disk) and 34 drives when reading the inner disk zones (36 GBps/disk).

Using a Dual Xeon computer with two Highpoint + one 3ware SATA controller and 24 disks we achieved 625 MBps read 534 MBps write. We observe that the Highpoint controllers show good throughput but behave poorly when more than one is present. This was a borrowed system and we did not have much latitude in reconfiguring it and explore this issue further.

To get to 1GBps, we built a white-box dual processor Opteron on a Tyan main board which includes one AMD PCI-X Bridge supporting 4 PCI-X slots. We added SuperMicro Marvell-based SATA controllers, as Brent Kelley of AMD reported great performance on these. Each of these controllers reliably delivers about 450 MBps sequential read and write with eight disks attached. The sequential disk read/write bandwidth scales linearly when a second SuperMicro card is added; but, with 3 of these cards and nineteen or more disks the bandwidth plateaus at around 1.05GBps read and 1.10 GBps write, using about 27% of one processor.

To get beyond 1 GBps we have been experimenting with the Newisys™ 4300. It supports up to four Opteron processors and includes three AMD-8131 PCI-X Bridges supporting four 64/133 PCI-X slots. We've tested the 4300 server with up to 48 disks, and the observed disk bandwidth scales almost linearly up to 32 disks (with 8 disks each on 4 SuperMicro SATA controllers), achieving a speed of 1.3 GBps with 24 disks and 1.7 GBps with 32 disks. To go beyond 32 disks, the slower PCI-X slots had to be used for the additional SATA controllers and bandwidth increased more slowly. The highest throughput (using the file system to a single logical volume) we've been able to measure on the Newisys™ server has been around 2.2 GBps. Note that NTFS doesn't support striped volumes of more than 32 physical disks, so for the 32+ disk tests we created two volumes and accessed files on these volumes in parallel.

Accessing the individual disks in a JBOD configuration yields around 2.5GBps. This is about 15% better than the NTFS performance. This may be due to file placement (inner tracks are slower and the files may be on inner tracks,) or it may be due to some software bottleneck – we have not characterized why NTFS is slower. One, of course, expects to pay a price for the convenience of a reliable log-based file system such as NTFS but the overhead seems to vary quite a bit. This area warrants further study.

We also measured the disk I/O bandwidth of a 32 processor Itanium® 2 NEC® Express5800/1320Xd running Microsoft® Windows® Server 2003 Datacenter Edition. An NTFS file volume striped across 21 Qlogic HBAs fiber channel connected to 41 Eurologic SAN blocs and over 900 disks. Although this system was configured for TPC-C testing (may random IOs) it delivers very good sequential bandwidth as well. This delivered 3.5GBps of disk IO bandwidth when treated as a JBOD and 2.5GB of bandwidth when treated as an NTFS stripe.

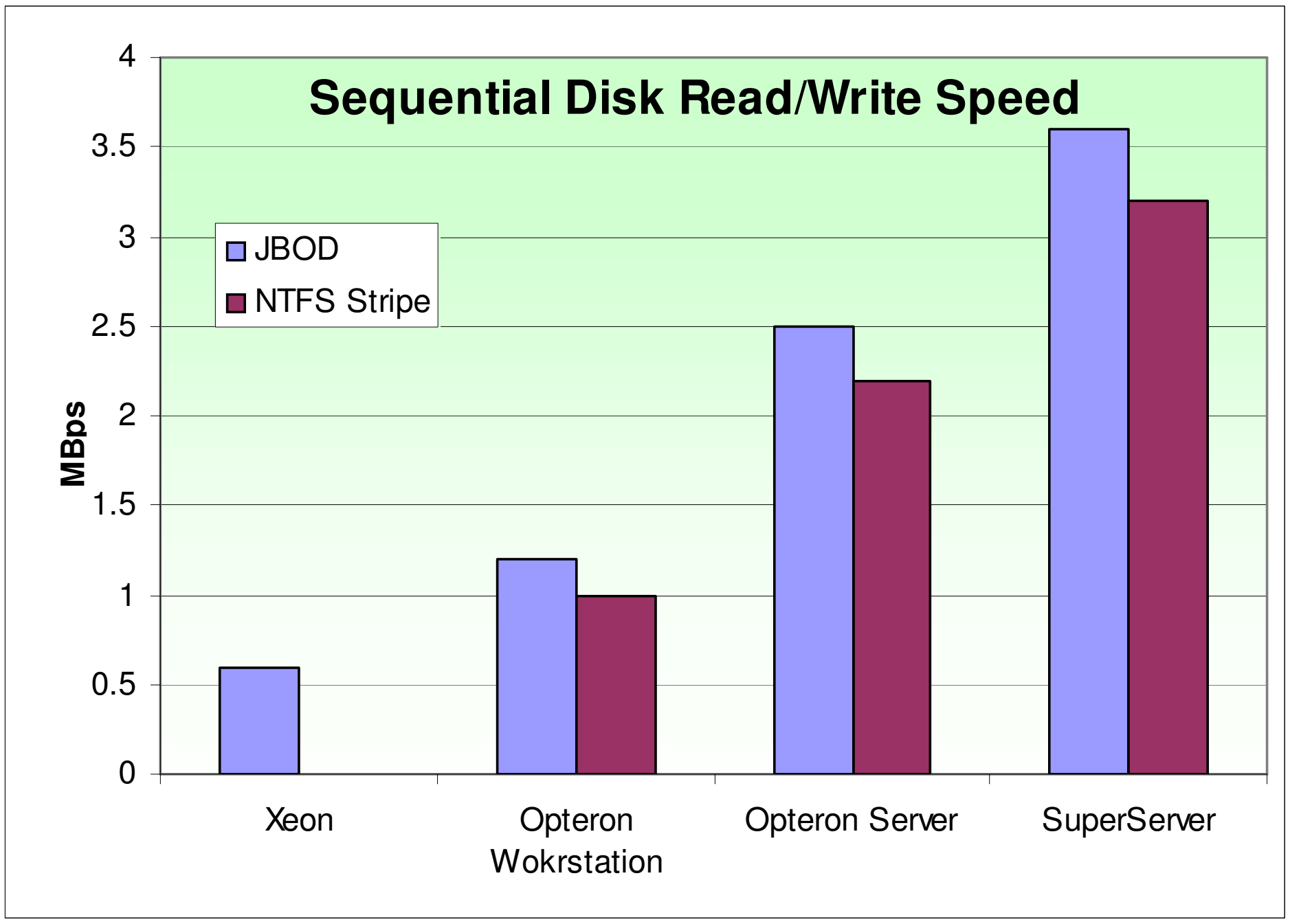

Figure1: Summary of sequential disk I/O bandwidth.

**Details – the Xeon/Highpoint/3ware configuration.**

The paper, *"A Quick Look at SATA Disk Performance,"* demonstrated 300MBps IO bandwidth using inexpensive computers. We repeated those measurements on a 24 disk array with a theoretical bandwidth of 1.4 GBps. The tested system was built to test the SqlServer2005 on the TerraServer workload. The test machine was a 5U Silicon Mechanics SM-524RX costing about 14k$:

- 2 x Xeon 2.4 GHz hyper threaded with.512 KB cache
- 533MBps front-side bus
- SuperMicro X5DPE-G2 motherboard
- 8 x 512MB of PCI 2100 ECC memory (4GB)
- 2 x HighPoint[1] Rocket Raid 1820 8-port SATA controllers (64bit 66Mhz)
- 1x 3Ware 8606-8 port SATA controller (64bit 66Mhz)
- 24 x Western Digital SE 250 GB 7200 rpm SATA disks
- Windows2003

This storage server is configured as 3 TB of hardware mirrored storage. Our first tests on that mirrored configuration created 2 GB files on each volume (using SortGen.Exe), the volumes were defragmenter, and then the following SQLIO test was done

```
sqlio -s30 -kW -fsequential -o4 -b1024 -BN -dfghijklm  temp.dat
```

Meaning it is a 30-second sequential IO test with 4 outstanding 1MB requests to the temp.dat file on volumes f, g,…., m. We varied the request depth, and number of volumes and of course tested both read and write. The other parameters were held constant.

The 3ware card had mirrored volumes F,G,H,I, while the 2 Highpoint controllers had volumes J,K,L,M and N,O,P,Q respectively. The throughput was disappointing (Figure 1) in part because the disks were not optimally laid out. Using 1MB 4-deep IOs, the 3ware card read at 137MBps and wrote at 138 MBps, the Highpoint card read at 221MBps and wrote at 176 MBps, The maximum sequential read rate using all the cards was 522MB MBps reading and 278 MBps writing. These graphs show the Sequential IO performance of the mirrored configuration. Drives O and P were slower because they were being used to host used for other data files.

---

[1] Highpoint controllers are relatively inexpensive and fast, but the Highpoint drivers have difficulty with more than one controller per system, and fail completely if more than two Highpoint controllers are installed. Hence we added a more expensive 3ware controller to mange the last 8 disks. Based on this experience, we moved to using SuperMicro SATA controllers that are both faster and less expensive than either Highpoint or 3ware.

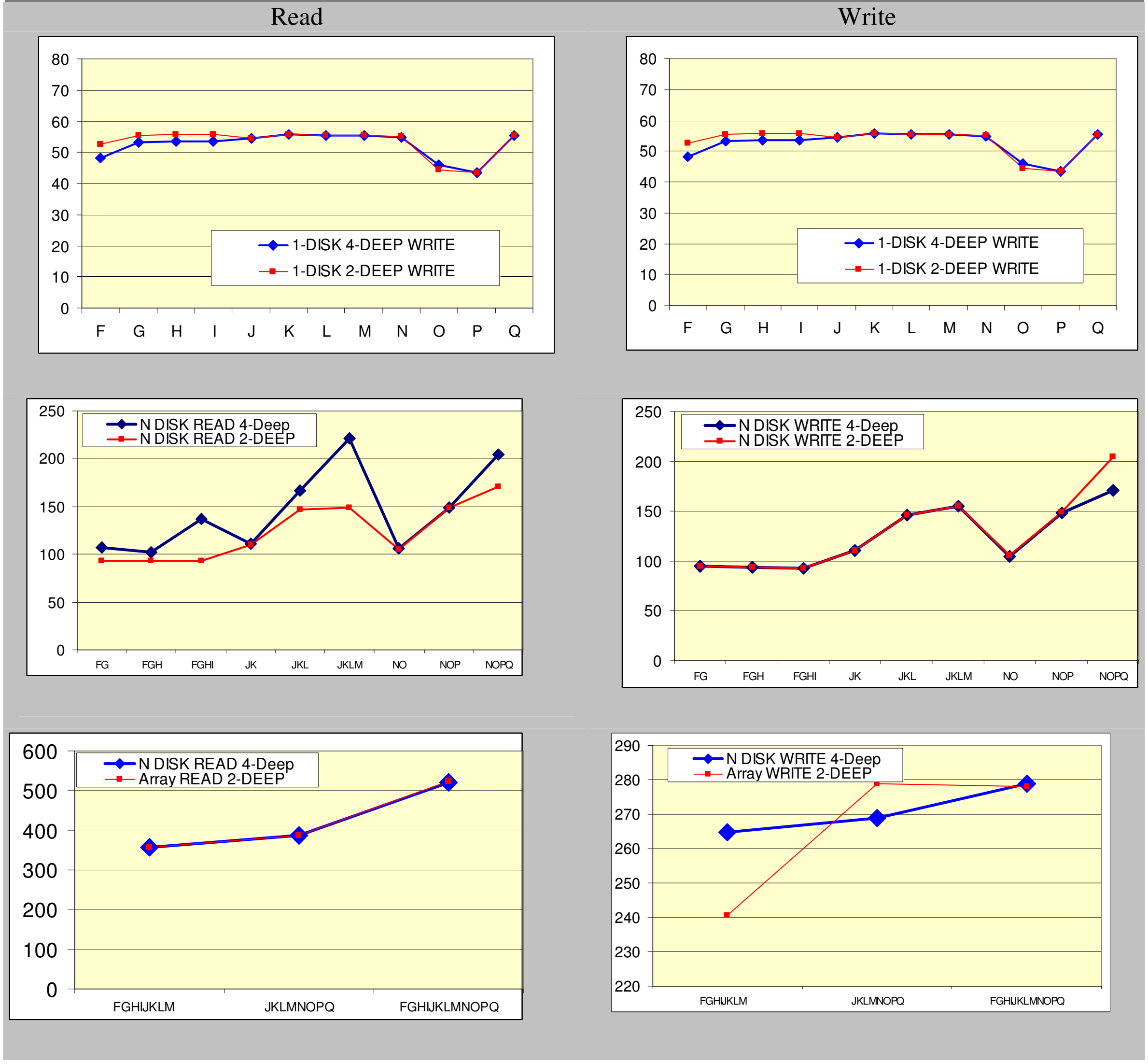

**Figure 1: Measurements on a Xeon Hardware Mirrored Configuration.**

To get a clearer performance picture, we reconfigured the system as Just a Bunch Of Disks (JBOD) spread across the three controllers. Figure 2 reports those experiments. Each disk reads and writes at about 60MBps. The 3ware controller saturates at 4 disks and 225MBps read and 200MBps write. Each Highpoint controller saturates at 6 drives with 346MBps read and 342MBps write. Combining two Highpoint controllers gives little more than the bandwidth of one controller. Still, the three combined controllers deliver 625MBps read rate and a 534MBps write rate on the full array. This is far short of the 1.4GBps capacity of the disks, but it is a very respectable IO rate. CPU load is an important issue. The third image in Figure 2 shows that on a dual 2.4 GHz Xeon reading at 600MBps, the operating system is consuming 10% of the CPU, about 20% of one of the two hyper-threaded Xeon processors. So there is 180% CPU capacity left to do networking.

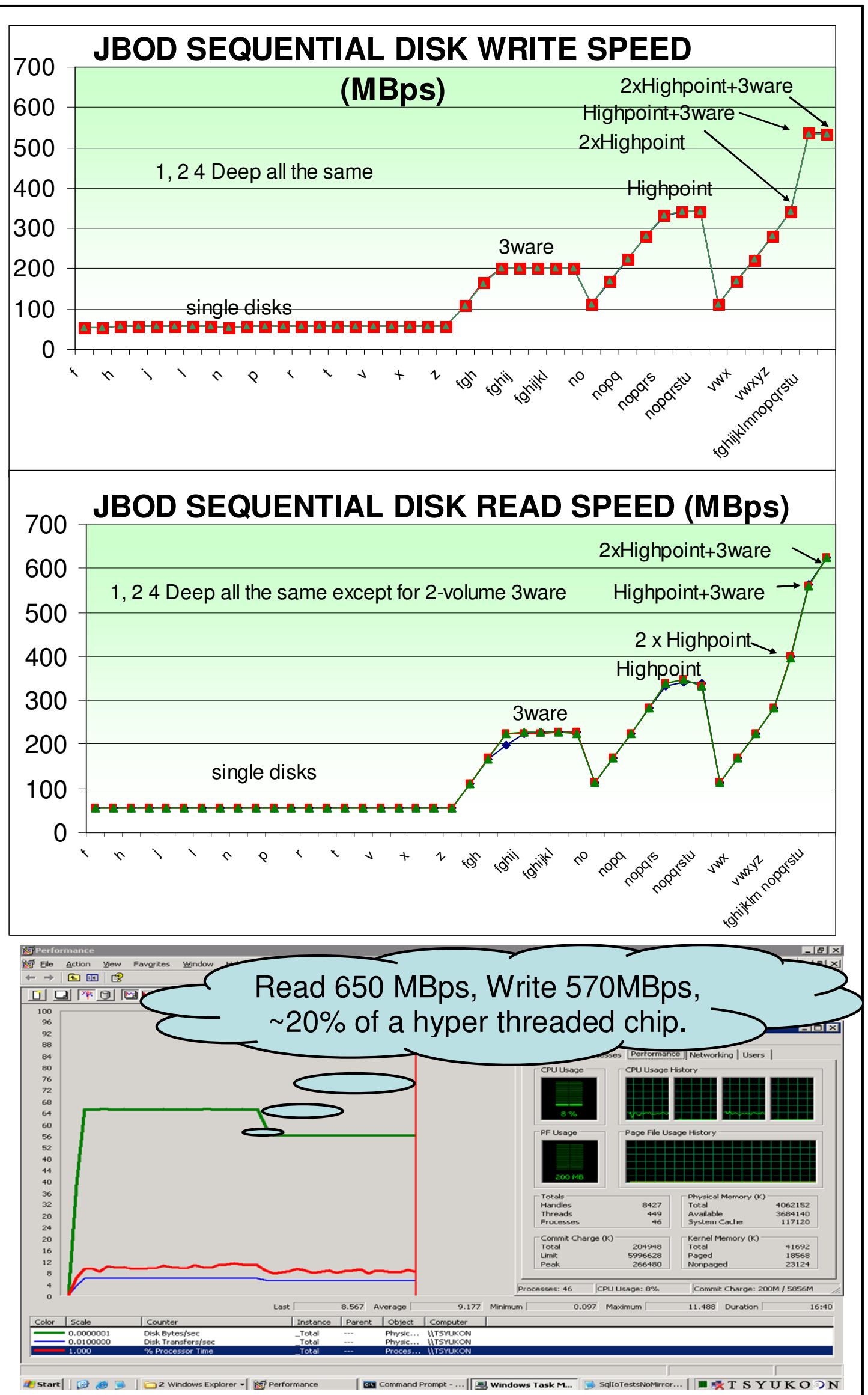

**Figure 2:** Xeon/JBOD sequential IO speed.

We conclude that the Xeon/Highpoint-3ware/Controller combination delivers 0.65 GBps sequential read and 0.55 GBps sequential write while using less half of a processor. This is a 14k$ computer.

We are exploring AMD white-box and Newisys™ solutions for inexpensive high-bandwidth storage bricks. The next section gives our preliminary and promising results.

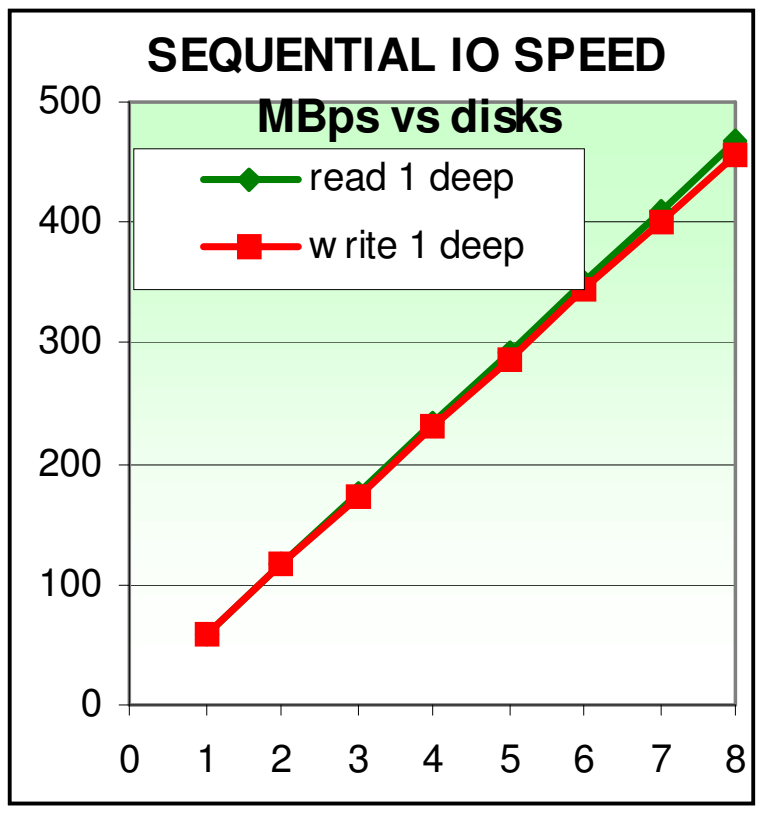

**Figure 3:** Unlike the other measured controllers (see Figure 2 at left) that plateaued at 4 or 6 drives, the Marvell-based SATA controller delivered the aggregate bandwidth of all 8 drives.

**The Tyan/Opteron/Marvell Configuration**

We measured Windows Server 2003 on a white-box system in 32 bit mode:

- Tyan S2882 Motherboard
- 2 x Opteron 246
- 4 x 1GB PC2700 ECC memory
- 3 x SuperMicro DAC SATA MV8 64-BIT PCI-X 8-port SATA controller
- 8 x 250GB WD Caviar SE WD2500JD hard drives
- 8 x 250GB Maxtor MaXLine Plus II hard drives
- 8 x 250GB Hitachi 13G0255 hard drives

The drives were configured as JBOD and tests were run on each individual drive and then scaled up 2, 3, …, 8 drives on one controller. The controller handles all 8 drives without plateau: see Figure 3. Then we measured 16 drives on 2 controller cards, and 20 drives on 3 cards. Adding drives beyond 20 appears not to increase disk IO bandwidth with the Tyan motherboard. With individual drives the speed is identical to the earlier tests - each drive reads and writes at a little under 60 MBps. With multiple disks the results show that the SuperMicro SATA controller (in combination with the Opteron box) is able to scale up to 8 disks without significant saturation, and throughput continues to scale almost linearly up to 16 drives on 2 controllers. Issuing 2-deep and 4-deep IO's did not speed things up, indeed, it sometimes slowed the system down (probably by confounding prefetch). See Figure 3.

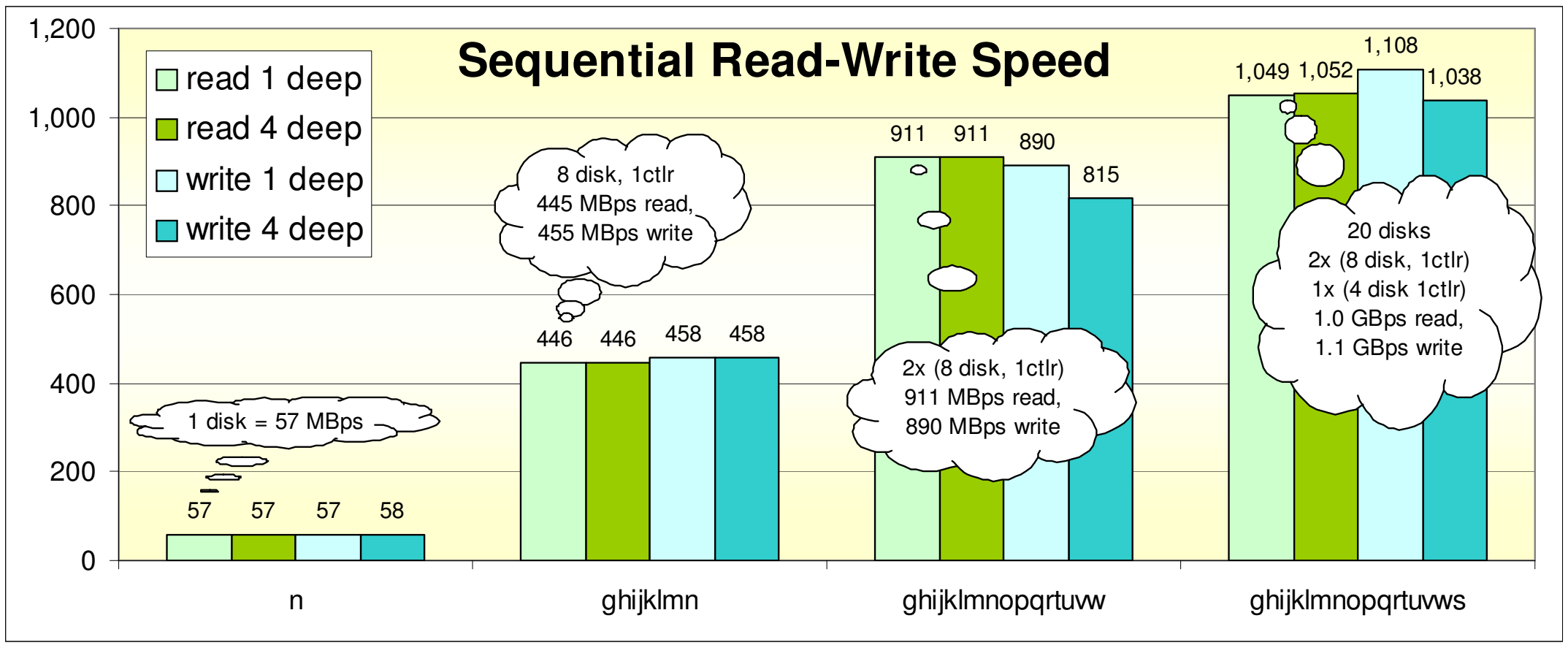

**Figure 3:** Opteron/SuperMicro Controller (Marvell® chip) performance.

Note that the speed at the high end may have been hampered somewhat by the mixture of disks from different manufacturers - we've observed small but consistent differences in speed between the Maxtor, Hitachi and WD drives; this merits further study.

We conclude that the single SuperMicro SATA controller is able to read and write at around 450 MBps with 8 drives. There is little benefit to queuing transfers more than one deep (the controller seems to be doing a good buffering job). The CPU utilization was 7% of one processor (3.5% per each processor) with 8 drives. Two controllers and 16 disks deliver almost exactly 2x better performance – about 900MBs, CPU usage goes up to about 23% of a processor (11% of each CPU). But, the write rate suffers a bit with 4-deep requests, and writes are slightly slower than reads. With 20 disks the throughput exceeds 1 GBps and CPU usage is around 27% of one processor (13% per processor).

**The Newisys™ server configuration**

The Newisys™ 4300 server supports four Opteron processors with up to 32 GB of DDR333 RAM and it includes three AMD-8131 PCI-X Bridges connected to four 64/133 PCI-X slots, two 64/100 PCI-X slots and one 64/66 PCI-X slot.

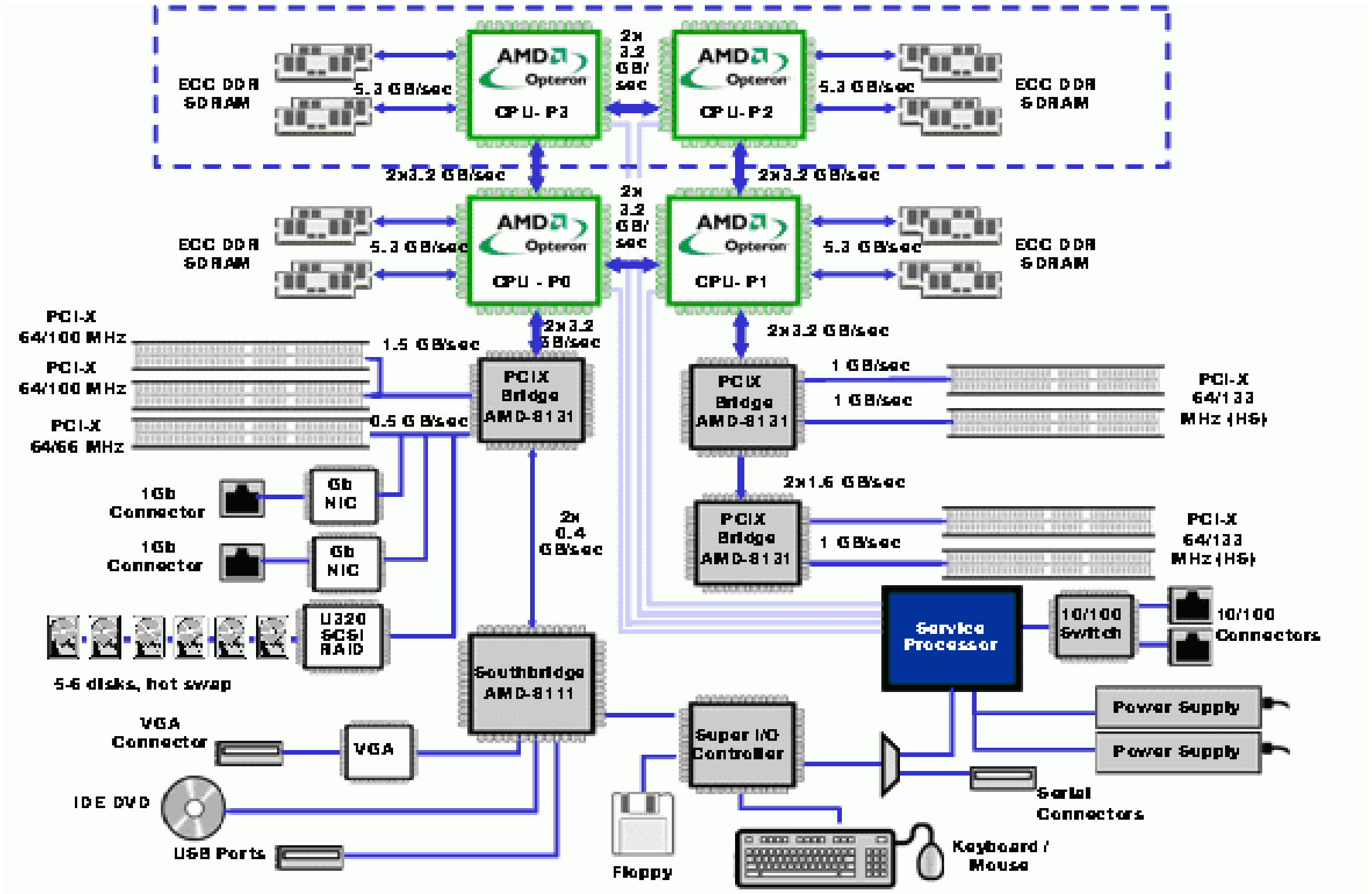

**Figure 4.** Block Diagram of Newisys™ 4300.

We ran a series of file I/O tests on this platform with up to 48 drives connected to six SuperMicro SATA controllers. We measured almost linear scaling up to 32 drives, and additional throughput above that point (though with sub-linear scaling); with all 48 disks the CPU utilization was around 50% of one CPU (i.e. 1/8th of the 4-CPU system).

For the tests that used 32 or fewer disks, we created a single NTFS striped volume (using 64K clusters ) that covered all of the disks; for 40 and 48 disks we created two NTFS volumes and placed a test file on each, these were accessed in parallel to effectively simulate a volume spanning more than 32 disks. The test file was created using the sortgen program (available for download at http://research.microsoft.com/~peteku/sortgen), with a size equal to the number of disks multiplied by 30GB. We measured the actual performance using diskspd (available at http://research.microsoft.com/~peteku/diskspd) with the following command-line parameters:

```
diskspd -x -h -d30 -o<count> -b<size> [-w] V:\test.dat
```

The –x parameter causes diskspd to report the results in a condensed format which is convenient for importing into a spreadsheet; the –h option disables all software buffering. The `-d30` option tells diskspd to run the test for 30 seconds (we also ran a subset of the tests for up to 10 minutes to make sure we were getting a reasonably accurate picture using the shorter tests, since it would be impractical to run all of them for a long period of time). The next two parameters vary in each test: `-o<count>` specifies the number of parallel / overlapped I/O requests to issue at a time (we used geometrically increasing values of 1 through 64), and `-b<size>` sets the size of the I/O buffer (it varied from 64 KB through 30 MB). The `-w` parameter was specified for the 'write' tests (and omitted for 'read' tests), and the last parameter indicates the file to use for the test.

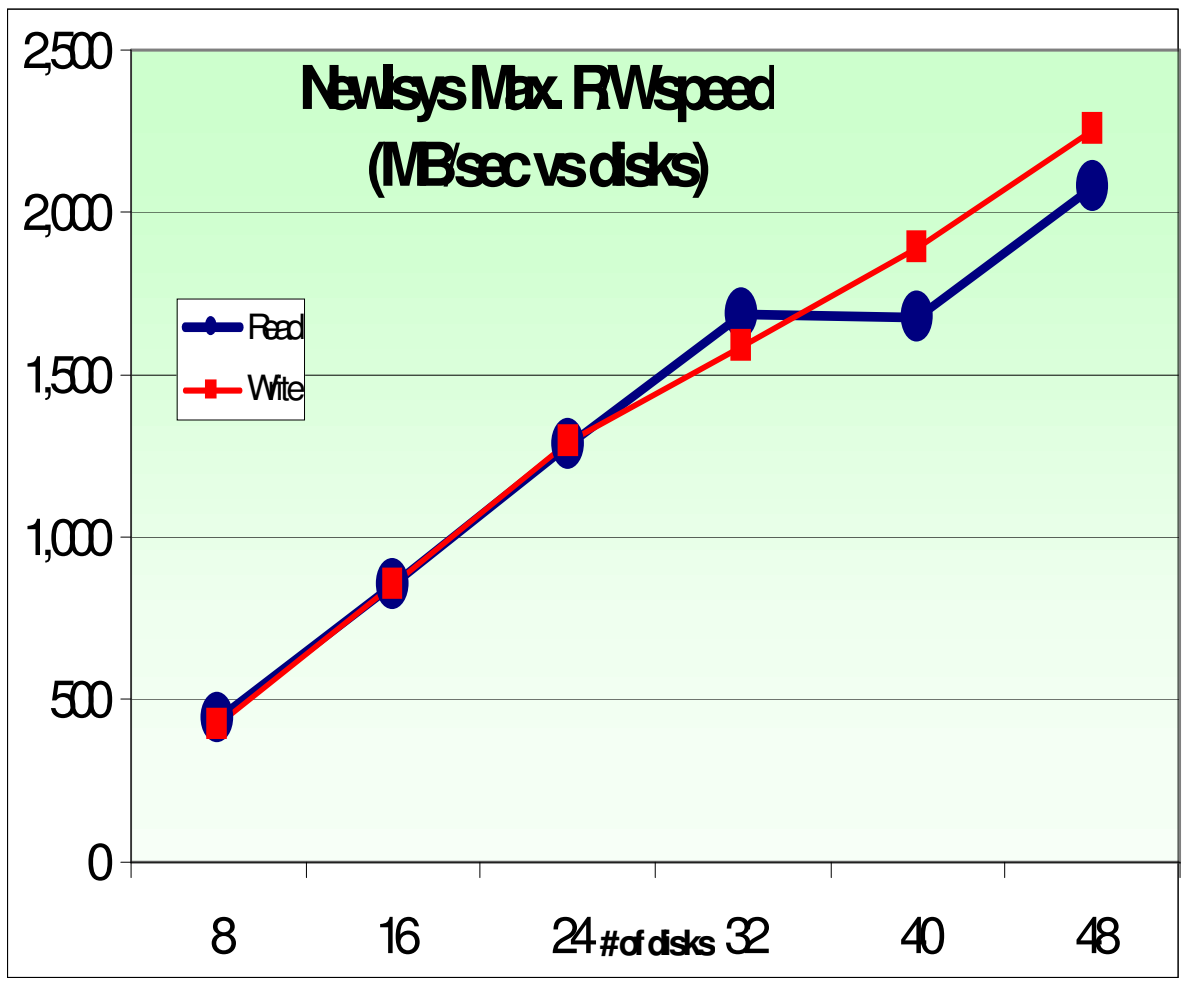

**Figure 5**. Sequential disk bandwidth on Newisys 4300.

The graph in Figure 5 shows the relationship between the number of disks and the read/write throughput. We see close to linear scaling, except for a little hiccup in read speed going from 32 to 40 disks; this is most likely due to two issues – using two NTFS volumes instead of one, and using slower PCI-X slots for some of the SATA controllers. There are a number of ways to configure the controllers and volumes for tests that involve more than 32 drives (since there are only four high-speed 64/133 PCI-X slots available and NTFS only supports 32 disks in a stripe set), and it's likely that further tuning could improve the performance.

The following series of graphs shows in detail how read/write bandwidth was affected by the number of overlapping I/O requests ("I/O depth") and block size for the various numbers of SATA disks attached to the Newisys™ server (the full data set is available in [XLS]).

Most of the graphs exhibit the shape one would expect – with increasing block size and I/O depth, the bandwidth quickly ramps up and reaches a fairly stead plateau. All of these tests were run with file system buffering disabled but the SATA controller and/or the hard drives were likely doing some degree of buffering (which may explain why writes usually appear slightly faster than reads).

Several of the graphs (for example, the 8-disk writes and both 48-disk tests) exhibit a rather unexpected unevenness - the maximum performance plateau looks more like a mountain range. We do not understand the cause of this. It is certainly the case that for maximum performance it is crucial to study the layout of the I/O connections and also test a number of configurations. We weren't able to reconfigure the servers enough times to delve into this issue more deeply.

As can be seen in the block diagram of the 4300 server (Figure 4. above), the slower PCI-X slots (on the left in the diagram) are connected to CPU #0 and all 3 connect to a single 8131 bridge; these are numbered 1,2,3 (going bottom up, i.e. the slowest 64/66 PIC-X slot on the bottom left is number 1). On the right-hand side, two of the high-speed 64/133 PCI-X slots are connected to the "lower" 8131 Bridge (these are numbered 4 and 5) and the other two 64/133 PCI-X slots are connected to the "upper" 8131 Bridge (and are numbered 6 and 7). All of the high-speed slots on the right-hand side are connected to CPU #1.

In the graphs below, we plot the speed as it relates to the I/O block size (which grows more or less logarithmically from 64K to 30M along the bottom left axis) and the number of outstanding (overlapped) I/O requests (from 1 to 64 along the bottom right axis). There are several patterns in the graphs that are readily apparent beyond the obvious rule that enough I/O's must be in progress to keep all the spindles busy. The writes generally reach the maximum speed plateau faster than reads, this is most likely because of write-caching at the firmware (controller) and hardware (disk drive) level.

The 8-disk test was run with a single SATA controller in slot 5, which is one of the high-speed PCI-X slots.

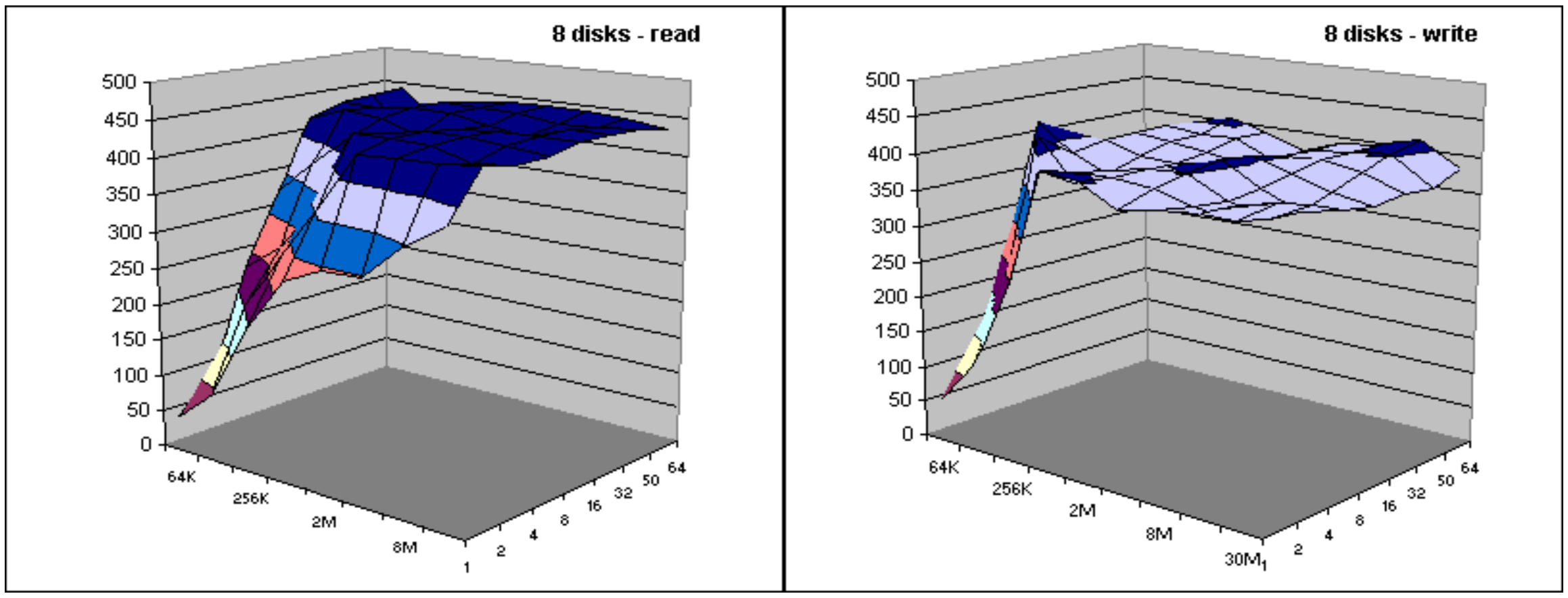

The 16-disk test used SATA controllers in slots 4 and 5, and the results look very much as one would expect:

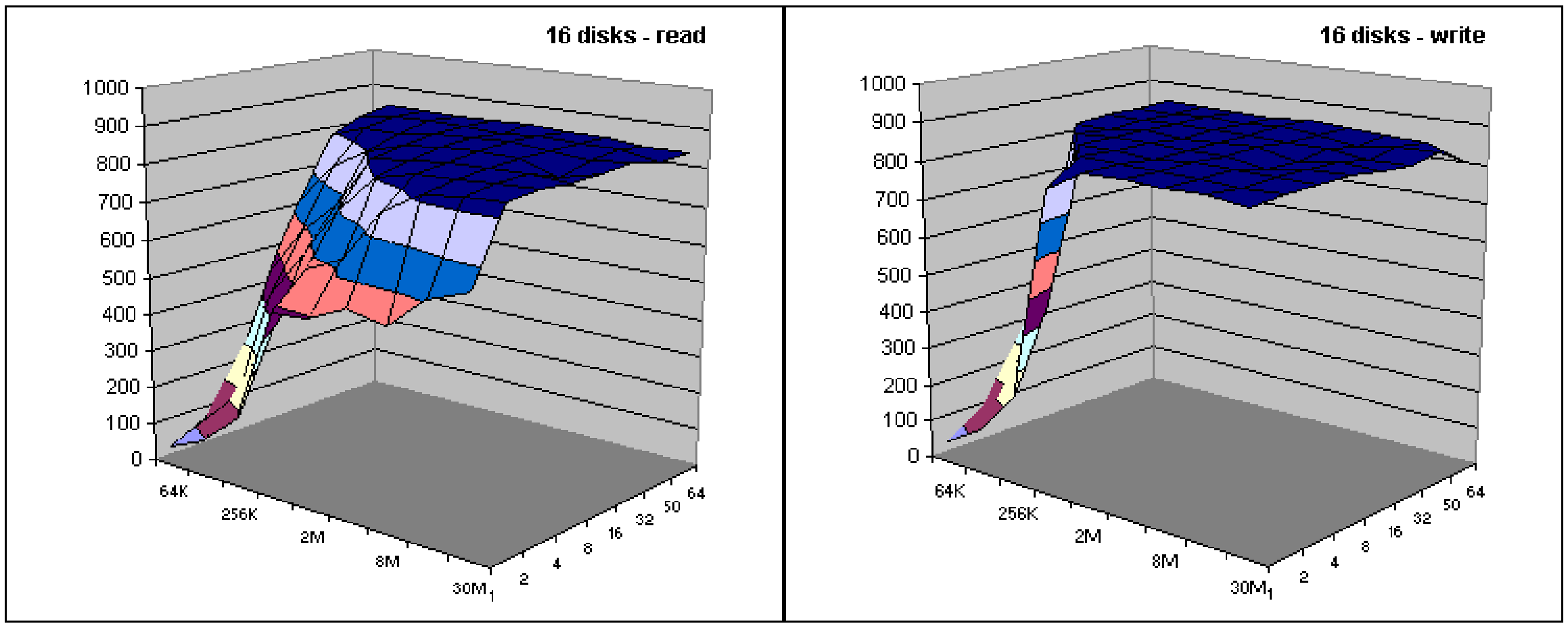

For the 24 disk test the 3 SATA controllers were in slots 4,5, and 6. Not surprisingly, it took quite a few outstanding I/O's and large blocks to reach maximum bandwidth with reads (the test file is now spread across 24 drives and it takes a lot of I/O to keep all the spindles busy); once again, the writes reach their plateau very quickly:

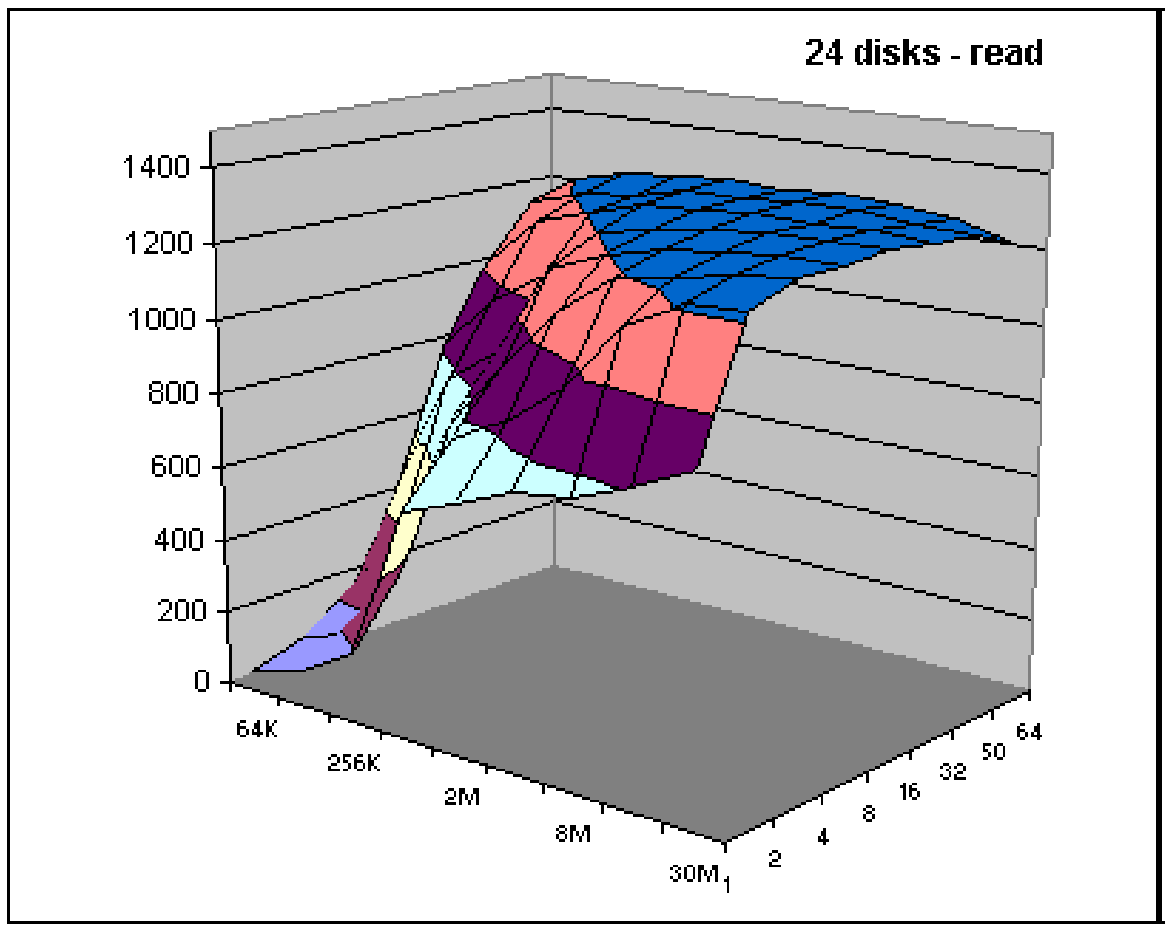

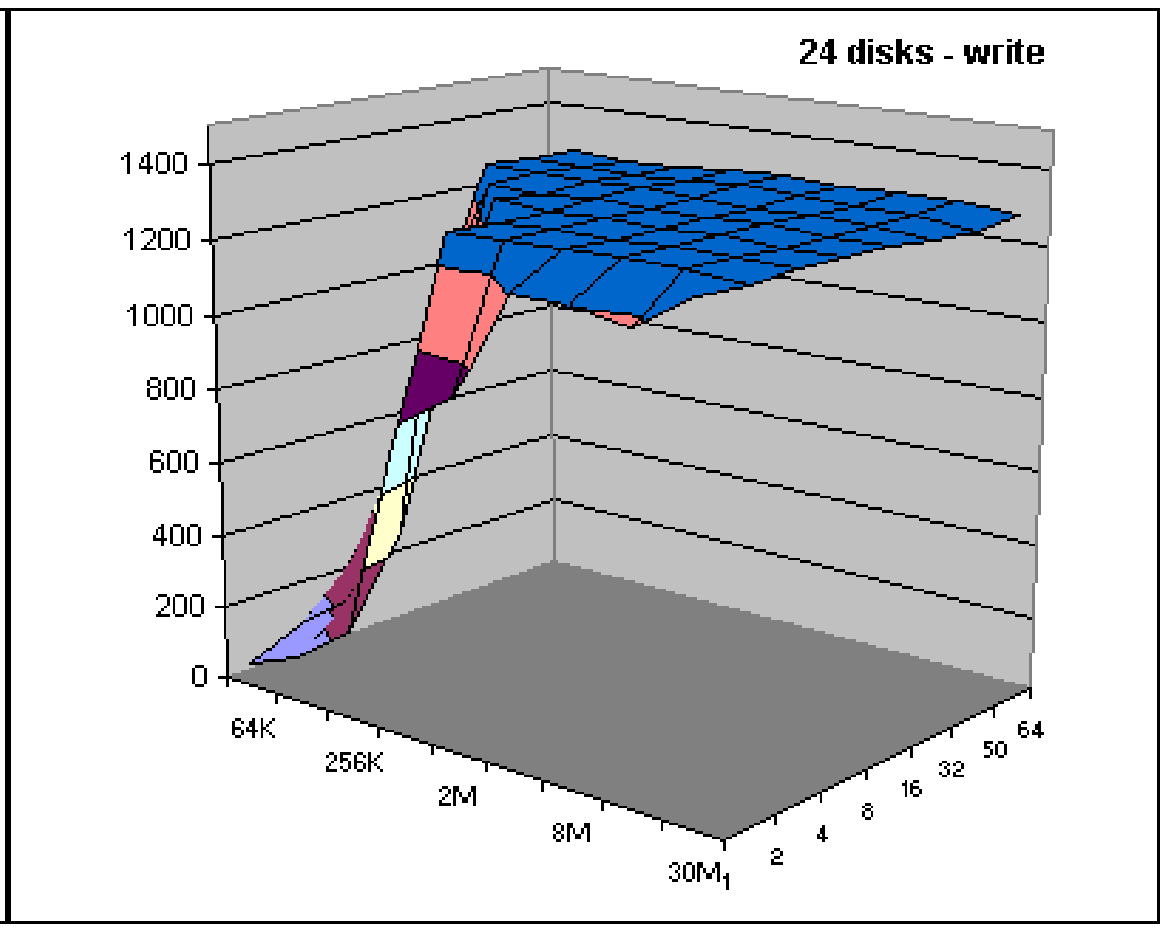

For the 32-disk the drives were connected to 4 SATA controllers in slots 4, 5, 6, and 7. The read graph looks as one might expect but the write graph shows a very uneven plateau; we're not sure what may have caused this:

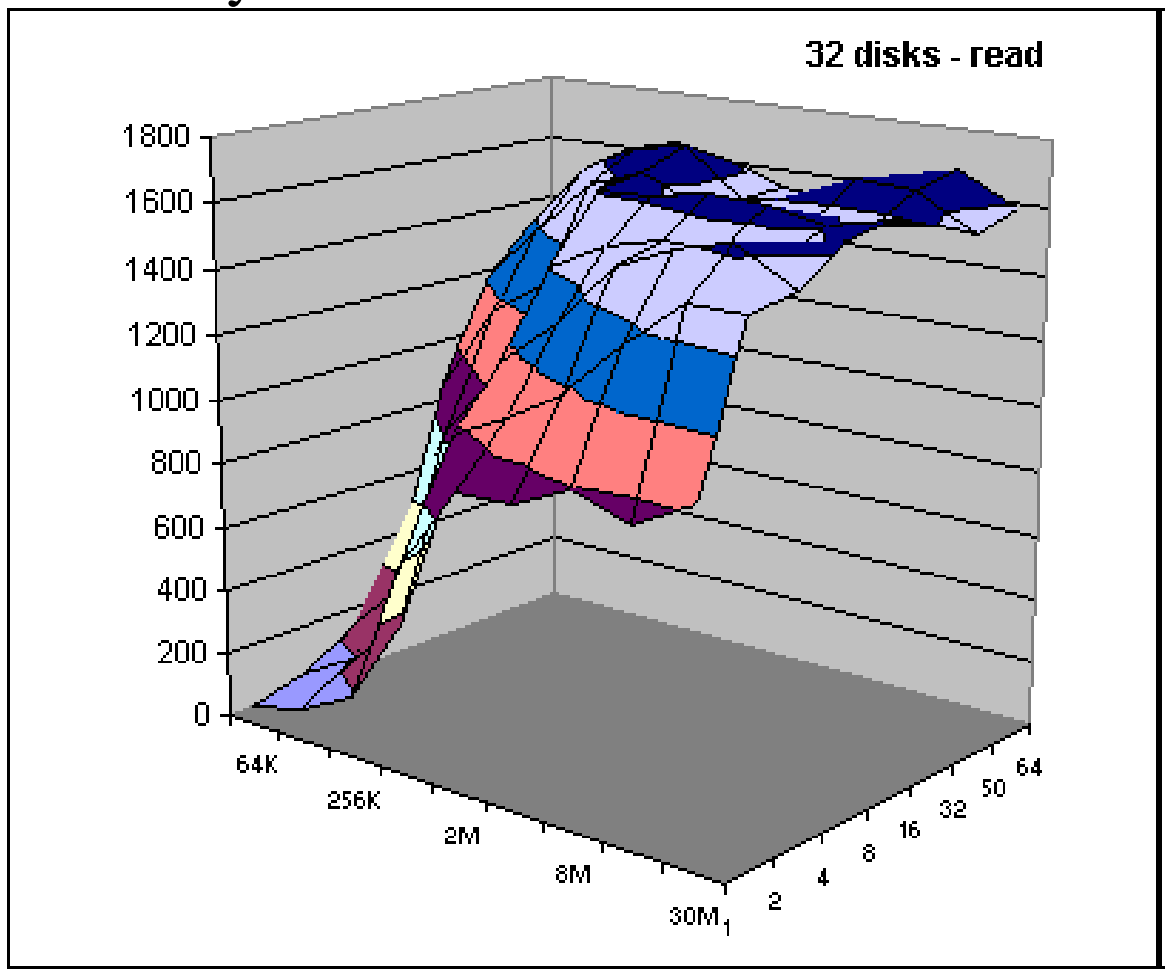

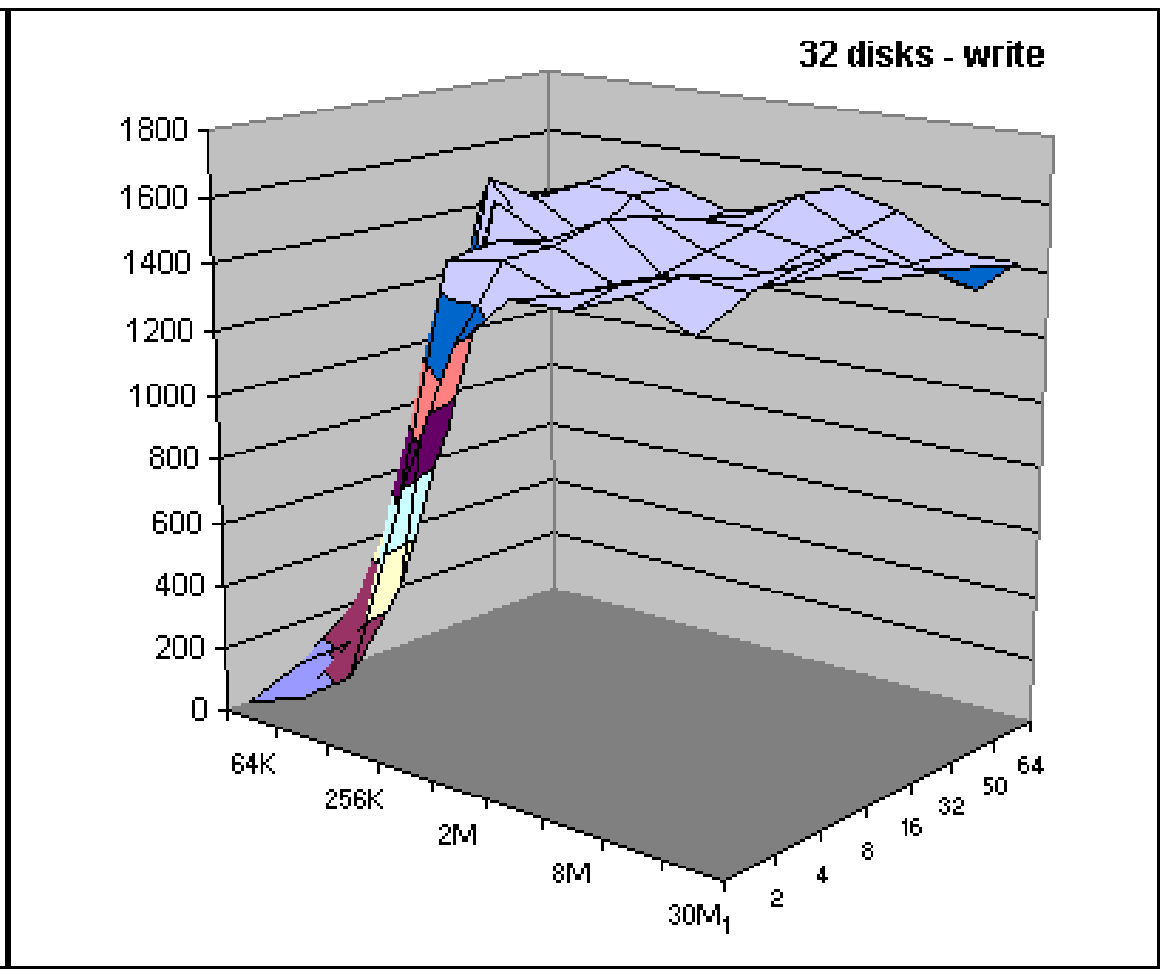

The read graph for the 40-disk case shows an unexpected "valley" in the top plateau. For this test we had to use one of the SATA controllers in a "slow" PCI-X slot (we're using 8-disk controllers and there are 4 high-speed PCI-X slots) and that may explain why the results aren't as clear-cut as with the tests using fewer disks. The writes show a very nice pattern, though, which may be explained by caching at the firmware / hardware level:

The highest number of disks we tested was 48 – we created two 24-disk NTFS volumes, one spanning drives connected to SATA controllers in slots 1, 4, and 5 and the other with disks on slots 3, 6, and 7; we then accessed test files on both volumes in parallel. The results look rather bumpy (esp. for the read tests), again this may be attributed to uneven speeds of the PCI-X slots being used:

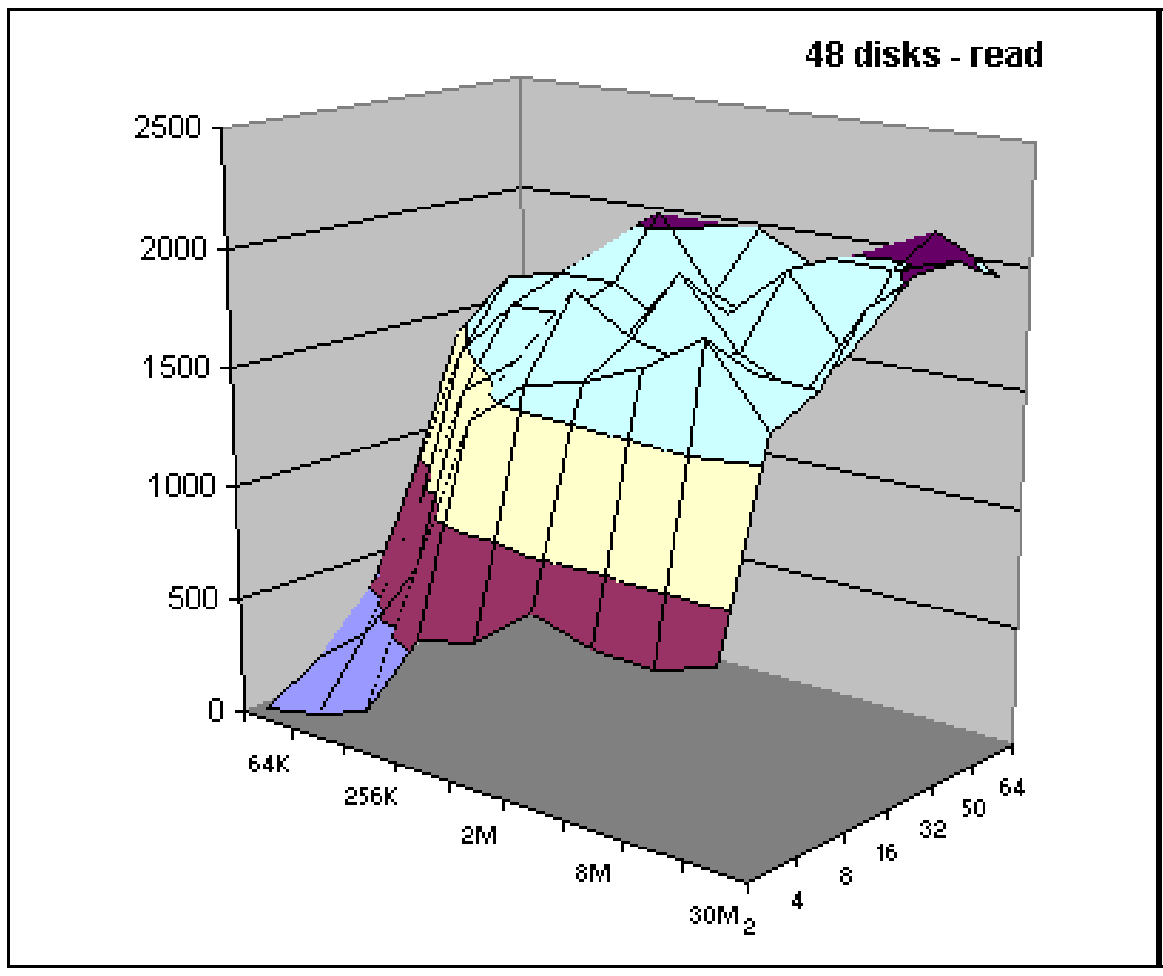

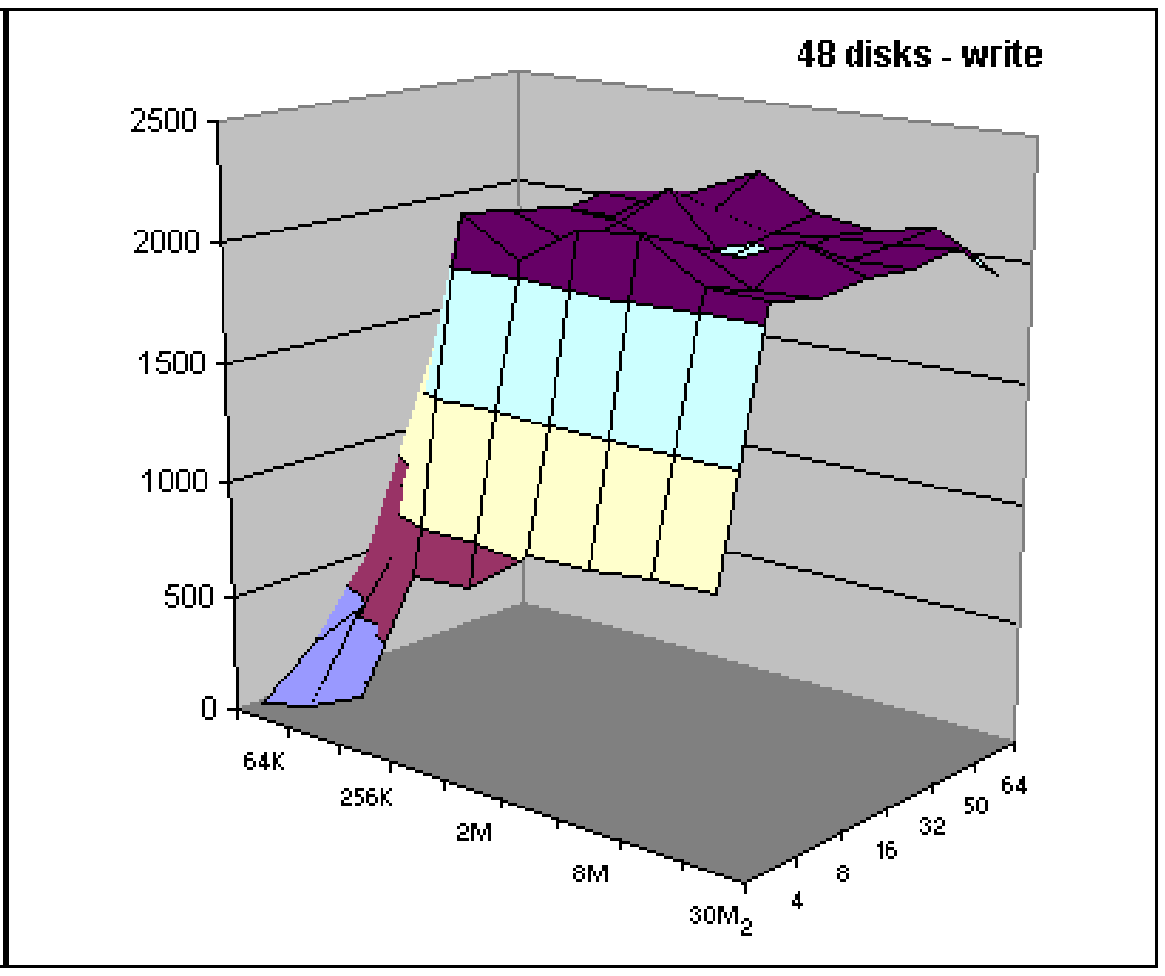

We conclude that the Newisys™ 4300 server, combined with the SuperMicro SATA controllers, is capable of delivering virtually linear file I/O scaling up to and including 32 drives (with NTFS file read/write speeds reaching about 1.7 GBps) and there appears to be plenty of additional I/O bandwidth available (with 48 disks we reached about 2.2 GBps). It is very important to place the SATA controllers in the right slots; using slower slot(s) for any of the controllers has a major impact on performance.

## The NEC 32 x Itanium2 and 20 Fiber Channel HBA Configuration.

With help from the SQLserver test lab and from Gerrit Saylor of Intel we experimented with a NEC® Express5800/1320Xd with 32 1.5 Ghz Itanium® 2 processors each with a 6MB L3 cache and 128 GB of main memory. The system has 904 fiber channel attached 36GB 15krpm SCSI disks connected via 21 Qlogic HBAs and 41 Eurologic SAN blocs. The system is running Microsoft Windows Server 2003 Datacenter Edition. Files reside on an NTFS striped logical volume.

Fiber channel runs at 2Gbps which translates to about 195 MBps of user data. Each HBA could deliver about 23,000 8KB IOps – so the system is well-balanced for a transaction processing workload. With 8 HBAs we were able to reliably read and write 1.5 GBps (very near the Fiber channel theoretical limit of 1.56GBps = 8*195MBps). At this rate we were using ½ of a processor when running a simple read application. An 8 HBA system was able to sort 34 GB in a minute and set a new MinuteSort record (reading and writing at more than 1GBps) [Nsort].

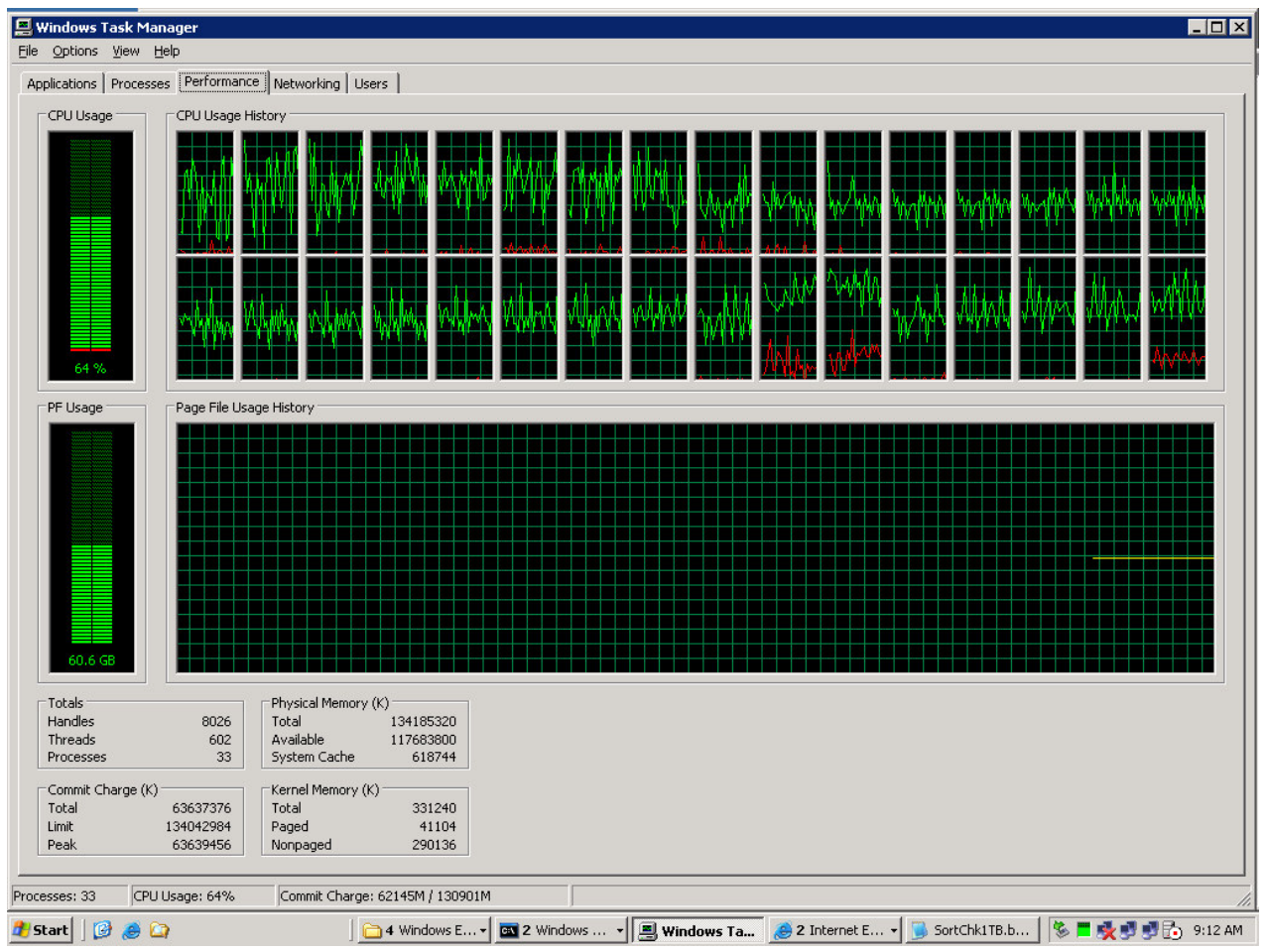

**Figure 6**: The NEC system is 2/3 busy during a terabyte sort.

In a subsequent test of the same NEC server with 20 HBAs, Nsort was able to sort a 1 TB file in 33 minutes – again reading and writing at about 1.2 GBps. The Nsort paper has the details. Figure 6 gives a sense that the processors are about 2/3 busy on this terabyte sort, the application is not being aggressive enough in using the IO system. It could probably sort the terabyte on this hardware in 22 minutes.

To explore the limits of the NEC system we measured a HBAs (each good for 195MBps). We used the SQLIO program to perform the tests, in each case using 4-deep 4MB requests (so 64 64KB SCSI IO requests were outstanding at any time). The resulting performance is shown in Figure 7. Clearly the hardware scales linearly. The “blue” JBOD (independent) lines show that the hardware scales nearly linearly. The machine has LOTS of IO bandwidth.

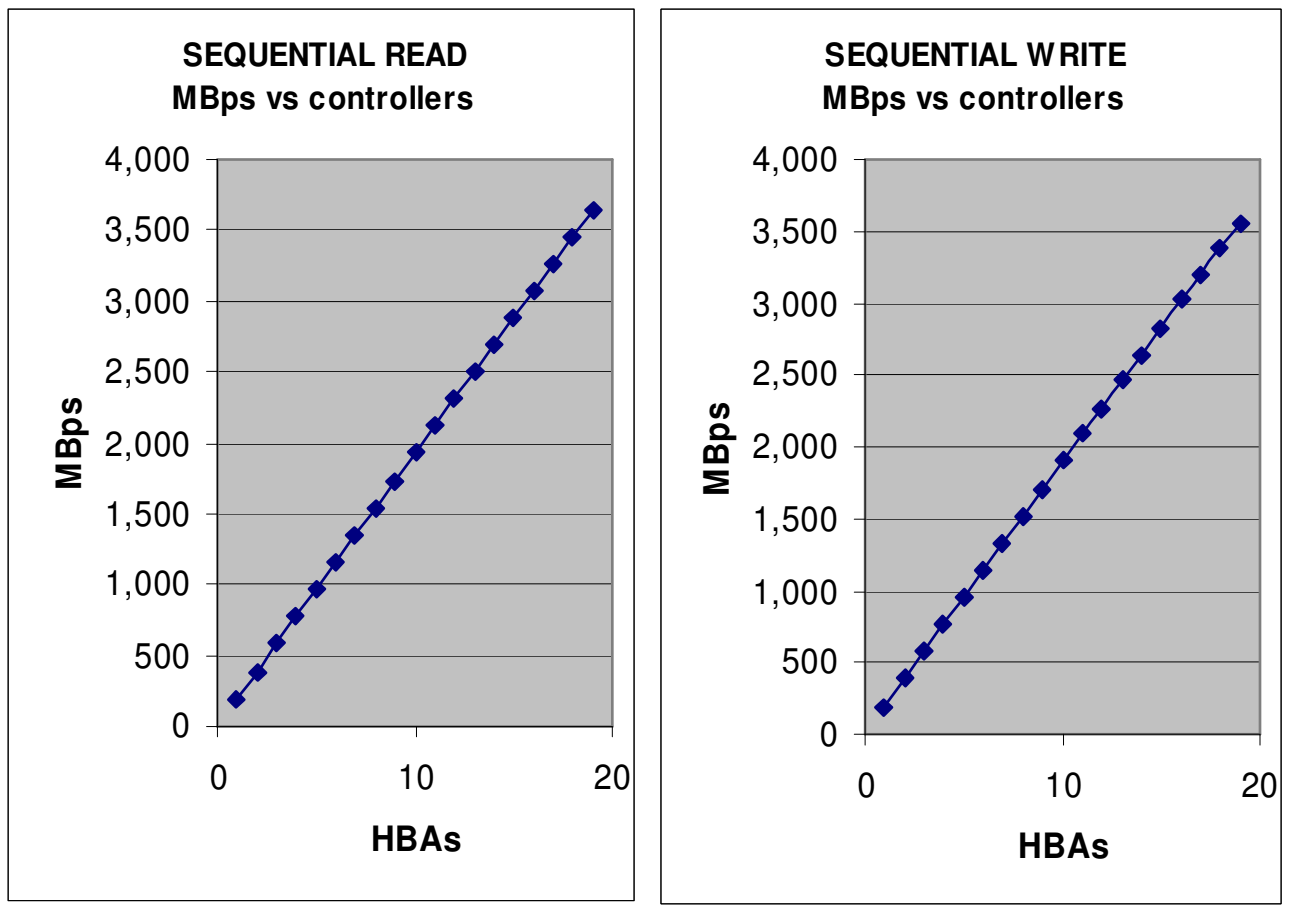

**Figure 7:** The NEC system delivers linear bandwidth as HBAs are added and treated as NTFS volumes.

**Acknowledgments**
Wyman Chung and Tom Barclay helped us set up the Xeon test system. The Newisys™ work was done jointly with Brent Kelly of AMD, and John Jenne, Dave Raddatz, Doug Norton, and Rich Oehler of Newisys™. Gerrit Saylor of Intel helped configure and tune the NEC Itanium 2 system. Brad Waters and Bruce Worthington of Microsoft gave us good advice on NTFS performance.